\documentclass{elsart}
\usepackage{epsfig}
\usepackage{amssymb} 
\usepackage{amsmath}
\begin{document}
\newcommand{\gk}{\vec{\gamma}\vec{k}}
\newcommand{\gE}{\gamma_0 E_k}
\newcommand{\ppl}{\vec{p}}
\newcommand{\bcm}{\vec{b}^{\star}}
\newcommand{\becm}{\vec{\beta}^{\star}}
\newcommand{\bepl}{\vec{\beta}}
\newcommand{\rcm}{\vec{r}^{\star}}
\newcommand{\rpl}{\vec{r}}
\newcommand{\A}{{$\mathcal A$}}
\newcommand{\wpk}{ \omega_{p-k}}
\newcommand{\Journal}[4]{ #1 {\bf #2} (#4) #3}
\newcommand{\NPA}{Nucl.\ Phys.\ A}
\newcommand{\PLB}{Phys.\ Lett.\ B}
\newcommand{\PRC}{Phys.\ Rev.\ C}
\newcommand{\ZPC}{Z.\ Phys.\ C}
\newcommand{\be}{\begin{equation}}
\newcommand{\ee}{\end{equation}}
\newcommand{\bea}{\begin{eqnarray}}
\newcommand{\eea}{\end{eqnarray}}
\begin{frontmatter}
\title{A new approach to solve the Boltzmann-Langevin equation for fermionic
 systems}
\author[catania,florence]{J. Rizzo,}
\author[france]{Ph. Chomaz,}
\author[catania]{M. Colonna}
\address[catania]{LNS-INFN, I-95123, Catania, Italy}
\address[florence]{Dipartimento di Fisica Universita' di Firenze, Florence, Italy}
\address[france]{GANIL(DSM-CEA/IN2P3-CNRS), F-14076 Caen, France}

\begin{abstract}
We present a new method to introduce phase-space fluctuations in 
transport theories, corresponding to a full implementation of the
Boltzmann-Langevin equation for fermionic systems.   
It is based on the procedure
originally developed by Bauer et al. for transport codes employing
the test particle method. In the new procedure, the Pauli principle is carefully checked, 
leading to a good reproduction of the correct fluctuations in
the 
``continuum limit'' ($h \rightarrow 0$). Accurate tests are carried out 
in one and two dimensional idealized systems, and finally results for a
full 3D application are shown. We stress the reliability of this method,
which can be easily plugged into existing tranport codes using test
particles, 
and its general applicability to systems characterized by instabilities,
like 
for instance multifragmentation processes.
\end{abstract}
\end{frontmatter}

\section{Introduction}
In recent years, the dynamics of heavy ion collisions at intermediate
energy has been extensively investigated within the framework of transport theories, such as the Nordheim approach, in which the Vlasov equation for the
one-body phase space density, $f({\bf r}, {\bf p},t)$, is supplemented with a Pauli-blocked
Boltzmann collision term \cite{Bertsch_rep,Aldo_rep}. 
The basic ingredients that enter the resulting transport 
equation, often called  Boltzmann-Uehling-Uhlenbeck (BUU) equation,  
are
the self-consistent mean-field potential and the two-body scattering cross sections. 
These transport models hence describe the time evolution of the reduced
one-body density in phase-space and, 
consequently, they are suited for
the description of one-body observables, such as inclusive particle spectra
in nuclear collisions, average collective flows and excitations.
However, they cannot provide a reliable description of fluctuation phenomena,
such as multi-fragmentation processes, i.e. the break-up of excited nuclear
systems into many pieces. In fact, 
neither fluctuations of one-body observables nor many-body correlations can be 
addressed with this class of mean-field models. 
Hence suitable extensions, including fluctuations of the one-body density, have to
be considered.

An intense theoretical work 
on fluctuations in nuclear dynamics has started in the past years, 
also stimulated by the availability of 
large amounts of experimental data on fragment formation in intermediate 
energy heavy ion collisions and the possibility to observe a liquid-gas 
phase transition \cite{Frankland,Moretto,Dagostino,Chomaz,EPJA,rep,Ono}.
In order to introduce fluctuations in transport theories, 
a number of different avenues have been taken, that can
be essentially reconducted to two different classes of models. 
One is the class of molecular dynamics models \cite{Ono,md,feld,ONOab,Pap,FFMD} while the other
kind is represented by stochastic mean-field approaches 
\cite{Ayik,Randrup,rep}.


In molecular dynamics models the many-body state is represented 
by a simple product wave function, with or without antisymmetrization.
The single particle wave functions are assumed to have a fixed Gaussian shape.
In this way, though nucleon wave functions are supposed to be
independent (mean-field approximation), the use of localized wave packets
induces many-body correlations both in mean-field propagation 
and hard two body scattering (collision integral), which 
is treated stochastically.
Hence this way to introduce many-body correlations 
and produce a trajectory branching is essentially based on the use of empirical gaussian
wave packets. 
If wave functions were allowed to assume any shape, the method would
become identical to standard mean-field descriptions. 
While the wave function localization appears appropriate
 to describe final fragmentation channels,  
where each single particle wave function should be localized within a
fragment,  
the use of fixed shape localized wave packets in the full dynamics 
could affect the correct description of  
one-body effects, such as spinodal instabilities and zero sound 
propagation \cite{FFMD,Akira_comparison}. 
 
On the other side, in the so-called stochastic mean-field approaches, 
the stochastic extension of the transport treatment for the one-particle
density is obtained by introducing a stochastic term representing the 
fluctuating part of the collision integral \cite{Ayik,Randrup}, in close analogy
with the Langevin equation for a brownian motion.
This can be derived as the next-order correction, in the equation describing
the time evolution of $f$, with respect to the standard 
average collision integral, leading to the Boltzmann-Langevin (BL) equation. 
Thus, the system is still described solely in terms of the reduced one-body density
$f$, but this function experiences a stochastic time evolution in response
to the random effect of the fluctuating collision term.
In this way density fluctuations are introduced, that are amplified when 
instabilities or bifurcations occur in the dynamics.  
This procedure is suitable also for addressing multifragmentation phenomena, 
since fragments can be associated with the regions where the spacial density
becomes larger, which finally can be reconstructed by sampling the 
one-body distribution function. 

A specific method for solving the Boltzmann-Langevin equation by direct 
numerical simulation was introduced in Refs.\cite{fluct_Chomaz,fluct_burgio}.
In this numerical implementation, the one-particle density 
$f({\bf r}, {\bf p})$ is
represented on a lattice of grid points in phase space and the collision integral is treated
by considering all possible transitions between phase space cells, adding a noise term whose
features are related to the average rate of transitions between two specified initial cells and
two final cells.  
The numerical implementation of this method has only been possible in two dimensions (2D) because it
requires too large computer resources in 3D. 

Hence several approximate solutions of the BL equations have been formulated,
mostly based on the projection of the BL noise only on a given dynamical variable
(such as the local quadrupole tensor of the momentum distribution) 
or on  ${\bf r}$  space only \cite{Suraud,Salvo}. 
More tractable fluctuating terms, such as a stochastic force added to the mean-field potential,
have also been proposed and extensively applied to 
multifragmentation studies \cite{BOB,EPJA,rep}.
 
However, the implementation of the full structure in phase space of the original BL term
can still be considered as an important goal to reach.
In fact, this would allow to treat a more general class of phenomena, where 
the correct description of fluctuations and correlations in  ${\bf p}$ space
is essential (such as particle production and fragment velocity correlations for instance).
Moreover, also in the multi-fragmentation mechanism, that is dominated 
by spacial density fluctuations, a more accurate representation of the 
full phase space dynamics, including fluctuations, would allow to improve the description of the fragment
kinematical properties and correlations. 

We also stress the general interest of this effort. Indeed transport phenomena occur in many physical
systems, for which a more precise description of the time evolution of the one-body distribution
function, including the effect of many-body correlations, 
would be important. 

A first attempt to introduce a fluctuating collision term in a 3D transport approach 
was made by Bauer et al. \cite{fluct_bauer}.
This method can be implemented relatively easily into standard transport codes that adopt  the 
scattering of pseudo-particles (or test particles) as a
method of solution of the collision integral  
and it consists essentially in forcing similar two-body collisions to occur
for neighboring test particles, defined according to a given distance in phase space, 
so that effectively two nucleons are involved in each particular
collision event. 
The distance should reproduce the phase-space shape of the nucleon wave function. 
In this way the random nature of the two-body scattering, that in the standard
codes applies only to test particles and is washed-out when using a huge number of
them, is transferred to entire nucleons. 
However, in the procedure proposed in Ref.\cite{fluct_bauer} the Pauli blocking is checked only for the collision of the
two original test particles and not for the entire swarm affected,
leading to some unpleasant features in the description of fermionic systems.  
Indeed
the Pauli-blocking violation introduces important inaccuracies in 
the fluctuations of  the one-body density. 

In the present manuscript we present a new method to reconstruct the phase space nucleon wave function in mean-field approaches,
in such a way that the Pauli-blocking is checked for the entire cloud of moved particles. 
This will improve the description of the 
fluctuation variance, 
approaching the one 
expected for fermionic systems.
We will also pay special attention to the definition of the phase-space metric that would
optimize the value of this variance.

The paper is organized as follows:
We will first recall the main ingredients of the BL theory, in order to connect
the formalism with the numerical implementation adopted (Section 2).
Then we will discuss in more detail 
the methods that have been proposed so far to solve the
Boltzmann-Langevin equation (Section 3).  
The new procedure that we follow to build fluctuations is presented in
Section 4. Several results demonstrating and analyzing in detail the method
are discussed from Section 5 to 8.
Conclusions and perspectives are drawn in Section 9.

\section{The Boltzmann-Langevin equation}

Within the semi-classical framework,
the stochastic transport equation of motion 
for the one-body distribution function $f$
can be expressed in the following form, 
\begin{equation}\label{BLE}
\dot{f}\ \equiv \ {\partial\over\partial t}f\ -\ \{h[f],f\}\ 
=\ K[f]\ =\ \bar{K}[f]\ +\ \delta K[f]\ ,
\end{equation}
where the left side describes the collisionless propagation
of the individual particles in their common self-consistent one-body field,
while the right side expresses the effect of the residual binary collisions.

The collision term $K[f]$ 
has a stochastic character.
For example, the distance a particle travels in the medium
before colliding is stochastic,
as is the resulting scattering angle.
At the Boltzmann level, 
Eq.(\ref{BLE}) includes only
the {\em average} part of the collision term, $\bar{K}[f]$. 
Usually, in nuclear systems,
the quantum statistics is taken into account
by adding suitable Fermi blocking 
in the single-particle final states, leading to the Uehling-Uhlenbeck
collision term \cite{Nordheim1928,UUPR43}.
The resulting nuclear Boltzmann equation 
exists in many implementations that differ with respect to
both the physical input 
(such as the types of constituents included,
the form of their effective Hamiltonian,
and their differential interaction cross sections)
and the numerical methods employed
(whether of test particle or lattice type)
and various names have been employed in the literature, including
BUU (for Boltzmann-Uehling-Uhlenbeck),
VUU (for Vlasov-Uehling-Uhlenbeck),
Landau-Vlasov and BNV (for Boltzmann-Nordheim-Vlasov) \cite{Bertsch_rep,Aldo_rep,Bao_an,seb,Bao_IBUU}.
The  Boltzmann-Langevin treatment
includes also the {\em stochastic} part of the collision term, 
$\delta K[f]$
\cite{Ayik,Randrup}. 


 
In the simple physical
scenario where the residual interaction can be considered as binary
collisions that are well localized in space and time, the average part is
given by 
\cite{Nordheim1928,UUPR43}: 
\begin{equation}
\bar{K}({\bf r},{\bf p}_1)\ =\ {g}\sum_{234}W(12;34)\left[ \bar{f}%
_{1}\bar{f}_{2}f_{3}f_{4}-f_{1}f_{2}\bar{f}_{3}\bar{f}_{4}\right] \ ,
\end{equation}
where $f_{i}$ is a short-hand notation for $f({\bf r},{\bf p}_i,t)$
and $\bar{f}\equiv 1-f$ is the associated Fermi blocking factor.
$g$ is the degeneracy factor. 

The basic transition rate is simply related to the differential cross section
for the corresponding two-body scattering process, 
\begin{equation}
W(12;34)\ =\ v_{12}\left({d\sigma \over d\Omega }\right)_{12\to34}
\delta({\bf p}_1+{\bf p}_2-{\bf p}_3-{\bf p}_4)\ ,
\end{equation}
being $v_{12}\equiv |\bold{v}_1-\bold{v}_2|$ the relative velocity, 
and it thus has corresponding symmetry properties, 
\begin{equation}
W(12;34)\ =\ W(21;34)\ =\ W(34;12)\ .  \label{Wsym}
\end{equation}

Since it arises from the same elementary two-body processes, the stochastic
part of the collision term is fully determined by the basic transition rate
as well, as a manifestation of the fluctuation-dissipation theorem. 
With the collisions assumed to be local in space and time,
the correlation function for the fluctuating part of the collision term
 has the following form,
\begin{equation}\label{dKdK}
\prec \delta K({\bf r},{\bf p_1},t)\ \delta K({\bf r}',{\bf p_1}',t')\succ = 
C({\bf p_1},{\bf p_1}',{\bf r},t)\ \delta ({\bf r}-{\bf r}')\ \delta (t-t')\ ,
\end{equation}
where $\prec\cdot\succ$ denotes the average with respect to 
the ensemble of possible trajectories resulting from 
the current one-body density $f$. 
Furthermore, for elastic scattering,
the correlation kernel is given by \cite{Ayik}: 
\begin{equation}
C({\bf p_1},{\bf p_1}',{\bf r},t)=\delta 
_{11'}\sum_{234}W(12;34)F(12;34)  
\nonumber
\end{equation}
\begin{equation}
+ \sum_{34}\left[ W(11';34)F(11';34)-2W(13;1'4)F(13;1'4)\right] \ ,
\end{equation} 

with the short-hand notations $\delta_{11'}\equiv h^D\delta({\bf p}_1-{\bf p}_1')$ 
and $F(12;34)\equiv f_{1}f_{2}\bar{f}_{3}\bar{f}_{4}+%
\bar{f}_{1}\bar{f}_{2}f_{3}f_{4}$ 
($D$ is the dimension of the space considered). 
The symmetry properties (\ref{Wsym}) of the transition rate 
ensure that the following sum rules hold,
\begin{eqnarray}
\sum_1 C({\bf p}_1,{\bf p}_2,{\bf r},t)\phantom{{\bf p}_1} =\ \sum_2 C({\bf p}_1,{\bf p}_2,{\bf r},t)
\phantom{{\bf p}_1} =\ 0\ , \\
\sum_1 C({\bf p}_1,{\bf p}_2,{\bf r},t){\bf p}_1 =\ \sum_2 C({\bf p}_1,{\bf p}_2,{\bf r},t){\bf p}_2
=\ \bold{0}\ ,\\
\sum_1 C({\bf p}_1,{\bf p}_2,{\bf r},t)\epsilon _{1}\, =\ \sum_2 C({\bf p}_1,{\bf p}_2,{\bf r},t)
\epsilon_2\,\, =\ 0\ ,
\end{eqnarray}
where $\epsilon_i=p_i^2/2m$ is the kinetic energy for a specified momentum.
These sum rules express the fact that each of the elementary binary
collisions conserves particle number, momentum, and energy, respectively.





\section{Methods to solve the BL equation}

\subsection{Lattice calculations}
The numerical implementation of the Boltzmann-Langevin equation 
 is accomplished by 
correctly simulating the basic stochastic process, i.e. the 
stochastic transition 
rate among phase-space cells. 
One can realize that this point 
is very delicate in finite systems, when we work with a relatively small 
number of nucleons, but still need to build a smooth distribution function. 
There are different ways to overcome this difficulty.

As mentioned in the Introduction, one possibility, 
proposed in Ref.\cite{fluct_Chomaz}, is to solve the BL equation on a lattice;
phase space is therefore divided into a number of cells, each one having 
volume $\Delta s = \Delta \mathbf{r} \Delta \mathbf{p}$
($s$ denotes a point in phase space: $s \equiv ({\bf r},{\bf p})$).

Each transition 
involves four locations, and the collision integral arises from the sum 
of the transitions evaluated for all possible combinations of the cells: 
from this consideration one recognizes that the practicality of the method is 
limited by the huge computational effort required. 
In fact, only two-dimensional implementations exist \cite{fluct_Chomaz,fluct_burgio}.
Each transition
represents a basic stochastic process, and, following the BL theory, 
the actual number of such transitions in a time step $\Delta t$
is dispersed around the mean value 
according to a Poisson distribution, so its variance amounts to:
\begin{equation}\label{sigma_ni}
\sigma^2_{\nu}=\bar{\nu},
\end{equation}
where the average is given by:
$$
\bar{\nu}(12;34)=\frac{\Delta s_1}{h^D}\frac{\Delta s_2}{h^D}
\frac{\Delta s_3}{h^D}\frac{\Delta s_4}{h^D}f_1 f_2\bar{f}_{3}\bar{f}_{4}
W(12;34)
$$
\begin{equation}
\cdot \delta(\mathbf{r}_{1}-\mathbf{r}_{2})
\delta(\mathbf{r}_{1}-\mathbf{r}_{3)}\delta(\mathbf{r}_{2}-\mathbf{r}_{4)}\Delta t.
\end{equation}\label{ni_med}
\\
This is a compact form of the fluctuation-dissipation theorem, since 
it predicts that the fluctuations are simply related to the mean number 
of dissipative processes. 
Then, a noise term $\delta\nu(12;34)$ is added to the mean 
number $\bar{\nu}(12;34)$, and thus the actual number of transitions 
is a random number picked from a normal distribution, the center 
of which is the mean value, and where the width is given by Eq.(\ref{sigma_ni}). 
In refs. \cite{fluct_burgio,chapelle} this procedure is proven to yield 
correct results for a 2D system of fermions interacting through hard two-body scattering.  
Starting from a non-equilibrium situation (two touching Fermi spheres)
the method 
is successful in reproducing the expected 
fluctuations, preserving also the average trajectory of the system. 
In particular, 
 the fluctuations introduced build the expected 
statistical value for
the equilibrium one-body density variance:
\begin{equation}
\sigma^2_f(s) = \prec \delta f(s)\delta f(s) \succ = f_{eq}(s)
\bar{f}_{eq}(s) h^D/(g\Delta s),
\label{variance}
\end{equation}
evaluated considering phase-space cells of volume $\Delta s = h^D$, for which
the fluctuating transition rate is implemented. 
Here $f_{eq}(s)$ denotes the equilibrium value of the one-body distribution
function and $\delta f(s) = f(s) - \prec f(s) \succ $. 
Of course, Eq.(\ref{variance}) holds also for volumes larger than $h^D$
and fluctuations are scaled accordingly. 
Also the co-variance, i.e. the correlation between density 
fluctuations in two different phase-space points, $s$ and $s'$,
is well reproduced \cite{fluct_burgio}.

The same authors of Ref.\cite{fluct_burgio}
also verified that, for the same idealized system, 
but prepared at low spatial density, 
early fluctuations developed in momentum space are subsequently 
transmitted into density fluctuations and amplified by the nuclear mean field 
\cite{fluct_burgio1}, 
leading to large instabilities and a statistical population of fragments. 

\subsection{The pseudo-particle correlation method}

The method of Ref.\cite{fluct_Chomaz} was originally developed as a way to overcome 
the problems arising from the solution of transport equations with  
the test particle method, which we now discuss briefly. 
Within the test particle method, each nucleon is represented by a 
collection of $N_{test}$ test particles, that are propagated according
to the mean-field interaction and random two-body scattering.
The use of a large number of test particles allows 
to have a smooth distribution function and a good 
coverage of the phase space for the Pauli blocking. 
On the other side, since collisions are treated stochastically for the
single test particles, the dispersion around the average number of nucleon
collisions (Eq.(\ref{sigma_ni})) is automatically divided by $N_{test}$. 
Hence fluctuations 
which are introduced by the test particle algorithm can also be seen as the
correct ones expected from the BL approach, divided by the factor $N_{test}$.
Various paths have been followed to overcome this problem. For instance, 
one can choose $N_{test}=1$, i.e. work with whole nucleons 
\cite{md}, but in this case one has to solve the problem of the 
induced numerical errors on the smooth path of the dynamics, 
i.e. the mean-field and the Pauli-blocking factors. Alternatively, Bauer et al.
proposed a method to introduce a correlation between close particles in 
phase-space \cite{fluct_bauer}.

The method follows the idea, applied in extended Time-Dependent-Hartree-Fock 
(TDHF) calculations, 
 of evolving the single-particle density 
including a statistical mixing of Slater determinants \cite{tdhf}.
The jump from 
one configuration to the other is possible under the important assumption 
that any coherence between determinants can be ignored 
(decorrelation approximation).  The choice of the single-particle basis 
for the determinant is however somewhat arbitrary, apart from the requirement 
of momentum-energy conservation. These ideas can be 
translated into the semi-classical approximation using the test particle 
method. The mixing between Slater determinants is realized in the collision 
integral by means of the following procedure. 
\begin{enumerate}
\item First of all, the nucleon-nucleon (NN) cross 
section is reduced by a factor $N_{test}$. 

\item Then two test particles $i_1$ and $i_2$ are chosen as 
colliding partners, and will be moved from  their positions ${\bf p}_1$ 
and ${\bf p}_2$, to new positions ${\bf p}_3$ 
and ${\bf p}_4$, 
according to the corresponding transition probability
including the Pauli blocking of the final states 
\cite{Aldo_rep}.

\item If the two particles can collide, the 
scattering actually involves two ``clouds'' of neighbouring test particles, 
corresponding to two entire nucleons ($2\times N_{test}$ particles). 
The contiguity criterion is based on the following definition of 
phase-space distance: 
\begin{equation}\label{distance1}
d_{ij}^2=(\mathbf{p}_i-\mathbf{p}_j)^2/p_F^2 + 
(\mathbf{r}_i-\mathbf{r}_j)^2/R^2
\end{equation}
where $p_F$ is the Fermi momentum and $R$ is the radius of the 
considered system. 
(It should be noticed that  the choice of this phase-space metric
is rather arbitrary).

\item In order to ensure energy and momentum conservation, the final states 
are adjusted 
using the average momenta of the two clouds as initial states. 
\end{enumerate}

\begin{figure}[htb]
\centering
\includegraphics[scale=0.5]{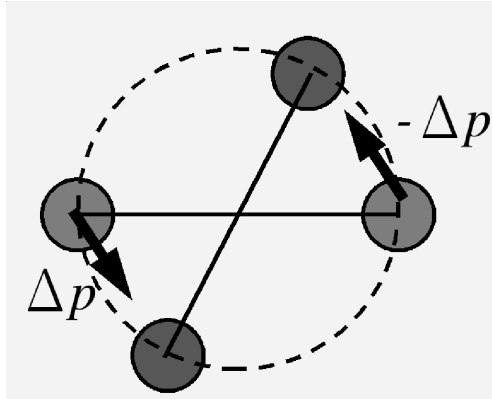}
\caption{Illustrative picture of nucleon-nucleon scattering in the 
pseudo-particle correlation method.}
\label{trasla}
\end{figure}

Two important remarks have to be done:
\begin{itemize}
\item the Pauli blocking is checked only for the collision of the test 
particles $i_1$ and $i_2$ and not for each particle of the cloud.  
We will see in the following that this 
choice induces a strong violation of the Pauli principle;  
\item the effect of the collision is a mere translation (in momentum space) 
of the two clouds to final positions, as sketched in Fig. \ref{trasla}. 
\end{itemize}

This method is conceptually simple, and moreover the implementation into 
existing BUU or BNV codes is really straightforward. Indeed it 
was immediately used to test the effects of the fluctuations on fragment 
mass spectra in a $^{20}Ne+ \,^{20}Ne$ reaction 
at $100\,AMeV$ \cite{fluct_bauer}. 

Its validity has been checked later by Chapelle et al. \cite{chapelle},
for the same idealized 2D system for which  
the BL lattice simulation method was implemented. 
Accordingly, a different metric,  
involving only distances in momentum space,  
was  used. 
As shown in \cite{chapelle},
the final equilibrium value obtained for the fluctuation 
variance does not reproduce the expected one. 
The variance profile, as a function of the energy,  appears to be proportional to $f_{eq}$ rather than 
to $f_{eq}\bar{f}_{eq}$. 
Actually, this would be the result for a classical 
system, obeying to the Boltzmann statistics. 
This can be considered as a consequence of the fact that Pauli 
blocking is checked only for the first two colliding partners, and the 
system progressively evolves towards an equilibrium configuration which 
is more consistent with the one of a classical system. 
From this discussion it is evident that a proper treatment of the Pauli 
blocking is a fundamental requirement for any numerical implementation 
 of the  Boltzmann-Langevin theory for fermionic systems. 
However, apart from these problems, the pseudo-particle correlation method 
represents a simple and practical way to implement fluctuations into 
transport codes. 
Therefore, it is worthwhile to think about some improvements to 
make it more accurate. 
This is the aim of the present work, as we will discuss in the 
next Section. 

\section{The improved pseudo-particle correlation method}\label{improved}

The method devised by Bauer et al. is able to agitate the phase space 
function and 
build fluctuations, although their strength does not reproduce the
expected value for fermionic systems.  
However, a careful modification of the original 
procedure can considerably improve the results. 

\begin{figure}[htb]
\centering
\includegraphics[scale=0.35]{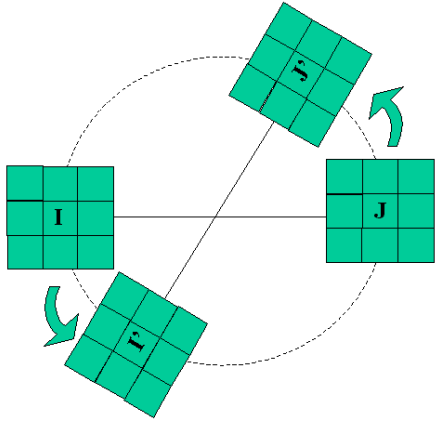}
\caption{Schematic picture of the collisional procedure in the improved pseudo-particle 
correlation method.}
\label{ruota}
\end{figure}

The new procedure can be summarized in the following steps:

\begin{enumerate}

\item The choice of the two colliding partners closely traces the 
standard recipe. If the two test particles are allowed to collide, 
two clouds of $N_{test}$ particles will be moved,
with conditions specified below.  

\item
 Only particles within a sphere, in coordinate
space,  
around the 
center of mass of the two partners $i_1$ and $i_2$ can belong to the 
clouds. The distance criterion is 
$|\mathbf{r}_i-\mathbf{r}_{CM}(i_1,i_2)|<d_r$, where $d_r$ is a free 
parameter (see later discussion).
Then, for 
this considered space sphere, 
 a grid is introduced in momentum space, 
around $i_1$ and $i_2$,
the size of each cell being $V_{cell}$. 

\item Given the momentum space cells $I$ and $J$ containing the partners $i_1$ and $i_2$, 
we consider the cells $I'$ and $J'$, corresponding to the final
positions,  in a rotated frame, 
as indicated in Fig. \ref{ruota}. For a given set of initial and final cells, 
the number of test particles that will be actually moved to final states is 
equal to the minimum between the occupation of the initial cells, $n$, and 
the availability of the final ones ${\bar n}' = (N_{max}-n')$:
$$
n_t(I,I';J,J')= \min(n_I,n_J,\bar{n}_{I'},\bar{n}_{J'}),
$$
where $N_{max} = g V_r V_{cell}/h^D $ is the maximum number of test particles that can stay
inside a cell. 
It should be noticed that this choice corresponds to the maximum of possible moves.

\item 
Surrounding momentum space cells 
are searched with the same prescriptions, until two entire nucleons 
are found. The search procedure is symmetric with respect to the 
center of momentum of the two partners
and random in the vicinity of the original cells (I,J). 

\item Finally, the two clouds are moved to the new states of the rotated frame
(see Fig.2),  so that 
the conservation of energy and momentum is automatically obtained. 
However, a further check is performed. 
In fact, due to the finite number of test particles, the one-body density is
not perfectly homogeneous inside each cell. 
This causes slight violations
of the conservations laws.  
Hence the origin of the momentum space grid is eventually slightly displaced in order to 
have a perfect energy and momentum conservation.

\end{enumerate}
This method involves two parameters, namely the radius of the sphere in 
$\mathbf{r}$ space, $d_r$, and the size of the momentum cells, $V_{cell}$. 
The radius $d_r$ fixes the spatial extension of the nucleon, that in turn 
influences its spreading in momentum space.
Hence this parameter fixes somehow the extension of the nucleon wave packet
in phase-space and could affect the transport 
dynamics.  
It can be constrained by physical 
arguments, and in general it should depend on the physical properties of the 
system, such as its dilution. 

\subsection{General comments}\label{comments}

The method proposed here 
has several advantages with respect to the original one of Ref.  
\cite{fluct_bauer}. It is 
still very flexible, and can be easily implemented into existing
transport codes; 
moreover, it is quite fast, and does not require huge amounts 
of resources.
The Pauli blocking 
is locally checked in each collision, so in principle this procedure 
satisfies the basic requirements to preserve the average evolution and 
to yield the expected fluctuations. Besides, differently from the original 
method, where the nucleon is essentially spherical in phase space, 
here it is allowed 
to assume any shape in 
momentum space, since the only restriction is given by the cell 
size $V_{cell}$. In 
other words, it is possible to take into account momentum
deformations of the nucleon 
wave packet, 
due to the momentum distribution of the system in particular conditions. 
For instance, when equilibrium 
is approached, 
the Pauli principle allows collisions essentially for nucleons
lying on the surface of the Fermi sphere;
therefore 
such nucleons should not have a spherical shape in momentum space.

It is also interesting 
to make a connection between the proposed method and the
lattice simulation method to build fluctuations. 
 
The numerical procedure described in Ref. \cite{fluct_Chomaz}, 
where the phase space is discretized by a fixed grid of 
cells,
allows to construct the correct value of the fluctuation amplitude 
in each  
cell of the grid, generally taken of volume $h^D$ 
(but in principle it 
can be even smaller), on the basis of the fluctuating gain and 
loss terms. 

On the other hand, our procedure is based on a different
philosophy: Instead of constructing fluctuations in fixed cells, 
when a collision occurs
we move all together the phase space volume containing
one nucleon.  The shape of this 
volume is not fixed a priori and depends on the location of the four
points (${\mathbf p}_1,{\mathbf p}_2,{\mathbf p}_3,{\mathbf p}_4$) 
that have been chosen as initial and final centroids of the considered 
two-body collision. This means that the ``correlation volume'', 
i.e. the volume of the sphere that envelops all test particles that move all 
together in each collision, is not fixed, but has to be considered as a dynamical
variable that generally exceeds $h^D$ and 
could even reach rather large values. 
In other words, 
compared to what is done in Ref. \cite{fluct_Chomaz}, we build 
fluctuations in volumes containing one nucleon, but these volumes 
do not correspond to single cells of a fixed grid: instead, their contour changes 
from one collision to another, according to the prescription adopted to 
construct the clouds around the two first colliding test particles 
$i_1$ and $i_2$.

But how can the two proposed methods be connected to the original BL equation ? 
It should be noticed that the Boltzmann-Langevin equation (\ref{BLE}) is a semi-classical equation,  
derived in the ``continuum limit'' ($h \rightarrow 0$).
This implies that the evolution 
of the system under consideration is completely determined by a smooth distribution function, 
i.e. it behaves as a fluid in phase space. 
In other words, the volume $\Delta s$ of the phase space cells for which
the fluctuating transition rate is evaluated, is supposed to be much larger 
than $h^D$. 
The Boltzmann-Langevin equation is
essentially based on the idea of a fluctuating collision rate among 
phase space cells
containing many nucleons. The formalism does
not provide any additional information about the ``structure'', i.e.
the shape of the single nucleon wave packet.

In both numerical implementations described above, 
two-body collisions among nucleons are treated as a stochastic process.
Only the definition of nucleon wave packet and/or of the elementary cells where
fluctuations are built change from one method to the other.
As discussed above, this 
ingredient is not contained in the BL equation and such information
is beyond its derivation. 
Hence, both procedures can be considered as correct 
implementations of the Boltzmann-Langevin theory in the ``continuum limit'',
the definition of ``nucleon wave packet'' being rather arbitrary.

However, 
it is also very interesting to investigate the differences between
the
various possible methods used to build the nucleon clouds, 
that in turn influence the correlation volume. 
In particular, one can also
try to force the
procedure in order to reduce the wave packet smearing and to correlate
particles inside smaller volumes (close to $h^D$).
 
\subsection{Details of the model}
In the present first implementation of the method, 
similarly to what was done for the lattice simulations of Ref.\cite{fluct_burgio}, 
we neglect 
the evolution in coordinate space, 
extending $d_r$ to the whole coordinate space. 
The cell size $V_{cell}$ is then constrained by the following arguments: it should be 
small enough to allow an accurate check of the Pauli blocking, but large 
enough to contain a sufficient number of test particles to reduce numerical 
uncertainties. 
In any case, $V_{cell}$  has not to exceed the volume where, at
most, one nucleon can be accommodated: 
$V_p = 
h^D/(g V_r)$,
being $V_r$ 
the volume 
in the coordinate space ($V_r =  \frac{4}{3} \pi d_r^3$).
In the following we take $g = 1$, unless expressly indicated.
Since we neglect the coordinate space, 
we will refer only to volumes $V$ in momentum space.
The expected equilibrium 
value of the variance $\sigma^2_f$ in such volumes is:
\begin{equation}
\label{scaling}
\sigma^2_f= f_{eq}\bar{f}_{eq}\frac{V_p}{V}=\frac{f_{eq}\bar{f_{eq}}}{N_{V}}
\end{equation}
where $N_{V}$ is the number of nucleons that can be accommodated, at most, 
in the considered volume $V$.  
It should be noticed that, 
according to our procedure to build the nucleon wave packet, 
the expected value of fluctuations, (Eq.\ref{scaling}),
can be reproduced only in volumes $V$ larger than 
the average
correlation volume, 
while it is underestimated
in smaller cells,
since fluctuations are built on a larger scale.

\subsection{Illustrative results}
In order to illustrate how our  procedure works, 
we will first consider the very simple case
of particles moving only along one direction. 
We take a large system, containing 1000 nucleons with average occupancy
$\prec f \succ = 0.5$.  So the total extension in momentum space 
of the system is $2000\times V_{p}$.
Then the system is divided into $N = 4000$ smaller cells that can contain at most
half nucleon, whose occupancy can be either $0$ or $1$. Hence $V_{cell} = V_p/2$.
The occupancy is randomly chosen at the beginning and collisions are performed
until equilibrium has been reached.
The initial ($p_1$,$p_2$) and final  ($p_3$,$p_4$) 
states of a collision 
are chosen randomly among the $N$ cells. We neglect energy and momentum conservation in this
simple example. 
 Moreover, for simplicity, we take ${p}_2 = -{p}_1 $ and  ${p}_4 = -{p}_3$.
Hence only half of the considered space is independent and, in this situation, 
the probability for a collision to happen is proportional to the product $f({p}_1)(1-f({p}_3))$.
According to Eq. (\ref{scaling}), the expected fluctuation value, in a given 
momentum space volume $V$, will be 
$\sigma^2_0 = 0.25~V_{p}/V$.  
The general prescription (see Sect. \ref{improved}) states that when a 
collision happens, a full nucleon has to collide. 
Therefore, we move together the cell $i$ and the closest occupied 
cell that finds an empty cell in the final position. 
This example can be seen as a simple implementation of the general 
procedure when considering nucleons represented by only two test 
particles.
One can easily realize that the  ``correlation radius'',
i.e. the distance between the two cells that move together, 
changes from one
event (collision) to another: sometimes the two co-moving cells are 
neighbours, sometimes they can be rather far. 
Of course, 
for practical reasons, we have to restrict our search within a given radius.   
For instance, considering 10 cells on each side (with
respect to $i$), we are able to reconstruct the nucleon and to 
perform the collision in more than 98$\%$ of the cases.   

\begin{figure}[htb]
\vspace{0.5cm}
\centering
\includegraphics[scale=0.25]{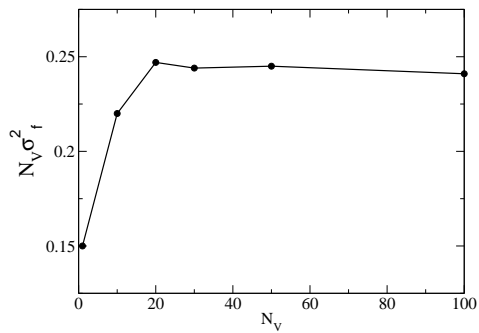}
\caption{Variance of the distribution function  
as a function of the number $N_V$ 
(see text),
in the idealized case of particles moving only along one direction.} 
\label{flu_joseph}
\end{figure}
In Fig. \ref{flu_joseph} we present the results for the  
fluctuation variance 
as a function of the number $N_V$ of ``nucleon'' cells 
(of volume $V_{p}$) that are contained inside the volume $V$ where the function $f$ is evaluated. 
One thousand events have been considered.
The fluctuation value presented in the Figure is multiplied by $N_V$.
As one can see in the Figure, 
the correct value of the fluctuations, i.e. $N_V\sigma^2_0 = 0.25$, is 
approximately recovered (within $\sim5\%$) for volumes 
$V \approx 20~V_{p}$.  
The slight deviation is due to the upper bound
considered for the  ``correlation radius'' and to the constraints
 imposed by the total mass conservation, that are also responsible for the
decrease of fluctuations observed 
for large $N_V$ values.
 
From this simple example, it already appears that 
the exact fluctuation value $(N_V\sigma^2_0 = 0.25)$ 
is reached only asymptotically.
This is due to the arbitrariness of the nucleon shape and of the 
centroid position.
Hence, when testing a given procedure to solve the stochastic 
collision integral, the resulting  
fluctuations have to be  
compared to the expected values
in large volumes, in which all different possibilities
(nucleon configurations) may be accommodated. 
 
These considerations would apply also to the original procedure
proposed by Bauer et al.  Thus, in Ref.\cite{chapelle}, the comparison
between the obtained fluctuation variances and the expected ones
should have been done considering also volumes containing many nucleons,
and not only cells of volume $V_p$
where, as seen in Fig.\ref{flu_joseph}, fluctuations are 
naturally suppressed by the fact that the nucleon wave packet may
take any shape.
However, this does not affect the conclusions drawn in \cite{chapelle},
mainly based on the wrong profile of the density variance
(and not on its absolute value). 
  
In the next Section we will move to describe more complex geometries
and different procedures to build the nucleon wave packets.

\section{Fluctuations of a Fermi gas: Analysis on the Fermi surface}

Here we will discuss our procedure to build fluctuations in the context of
a system of fermions having
a given density and temperature, i.e. initialized according to a Fermi-Dirac
distribution. This is a situation that is easily reached in the course
of a dissipative nuclear reaction, after the initial collisional 
shock \cite{rep}.   
Since collisions happen mostly on the Fermi surface, let us first 
consider only particles within a thin stripe $dp$ around the
Fermi momentum, where 
$\prec f \succ = 0.5$. 

For those colliding particles 
the relative velocity $v_{12}$ is constant, 
since only particles with opposite
momenta can collide, in order to stay on the Fermi surface. So 
the transition rate is essentially governed by   
the product
$f_1f_2\bar{f}_3\bar{f}_4$.

This simplified system will allow us to
study the sensitivity of the results to the
ingredients of the procedure used to build, step by step, 
the nucleon clouds.   
As also discussed before, 
we will show that 
the correct 
fluctuation amplitude is recovered in the ``continuum limit'', i.e. in large
volumes. 
However, in the following 
we will mostly concentrate 
on the fluctuation amplitude in volumes of 
size $V_{p}$, in order to probe  
the extension  
of the nucleon wave packet.  
The latter could influence significantly
the evolution of 
actual nuclear collisions, where it could be important to work with 
more compact nucleon configurations.

Using this simple model we will find that building the maximum fluctuation
value (e.g. $\sigma^2_0=0.25$) in such a volume, i.e. correlating
particles inside $V_p$, 
is a quite difficult task and additional efforts are required in order to 
reduce  the nucleon cloud smearing. 


\subsection{Details of the simplified model}
\label{details}
In order to keep the particles on the Fermi surface, 
the original procedure described in Sect. \ref{improved} was slightly 
modified so that the modulus of the momentum of the particles does not 
change: only particles belonging to cells with opposite momenta can collide. 
Then, for symmetry reasons it will result 
$f(\mathbf{p}_1)=f(\mathbf{p}_2 = \mathbf{-p}_1)$.
The 
transition probability finally depends only on two values of the distribution 
function, $f(\mathbf{p}_1)$ and  $f(\mathbf{p}_3)$. 
We use 
spherical coordinates $\big(\cos (\theta),\phi \big)$ 
as independent variables. The collision 
partners of particles in the cell $\big( \cos (\theta),\phi\big)$ 
will be the ones in the cell $\big( \cos (\pi-\theta),\pi+\phi\big)$. 
The choice of these coordinates implies the grids used in the procedure to be 
fixed along the $\cos(\theta)$ axis. They can slide along the $\phi$ axis, 
where periodic 
conditions 
are taken into account. 

The procedure is 
as follows: a test particle $i_1$ is chosen at random
(together with the corresponding partner $i_2$) and a grid is constructed, 
centered around $i_1$ and $i_2$ (see Fig.2).
The rotation angles, also taken from 
a flat distribution, fix the final states and the new frame
corresponding to these final positions.  

We consider a shell with $40\times40$ cells, whose size 
is chosen to accommodate one nucleon at most, 
so that
$V_{cell} 
=V_p$. 
We have checked that the results do not depend on the number of cells,
i.e. the number of nucleons considered, 
by performing the same calculations on a $64\times64$ system. 
We use $500$ test particles per nucleon.

The growth of fluctuations is described by the variance $\sigma^2_f$ 
of the distribution function, as discussed before.  
Usually it is evaluated performing an ensemble average over a large
number of events having the same initialization. However, in this scheme 
the mean value $\prec f\succ$ does not depend on the position ${\bf p}$ and  
is not changing in time.
Considering the large number of cells employed, we can also 
follow the evolution of the system by calculating in a single event 
the following quantity:
\begin{equation}\label{sigmaf}
\sigma^2_f=\langle(f_i-\langle f\rangle)^2\rangle _{cells},
\end{equation}
where $\langle\cdot\rangle $ denotes the average with respect to 
the cells and $\langle f\rangle = 0.5$.  


\subsection{``Fixed grid'' calculations}
\label{fixed}

In order to understand how the recognition of the nucleon wave packet can be influenced by 
the details of our procedure,  
we first further simplify the problem 
by using a single 
fixed grid to make the collisions, i.e. test particles are placed at (and can only move to) the center
of the cells of the grid. 
In this case we have a single reference 
frame, so the effect of a collision (in the space 
$\big(\cos(\theta),\phi\big)$ ) is a mere translation to other states of the same grid.
Apart from the use of test particles, and related differences in the method
followed to solve the collision integral, 
in this way the procedure becomes 
similar to the one used in Ref. \cite{fluct_Chomaz}. 

We will take, 
as initial conditions, $f$ strictly equal to $0.5$ 
in each cell of 
the fixed grid,  
corresponding to $800$ nucleons. 
By construction, 
once collisions are performed, the only possible values for $f$ are $0$, 
$0.5$ and $1$.

The maximum 
fluctuations in $V_p$ ($\sigma^2_0=0.25$) would be reached only if, in each cell, $f$
is either zero or one. 
Achieving this result is not  a trivial task: 
Cells starting with $f=1$ can be 
partially emptied, and cells with $f=0$ partially filled. 

\begin{figure}[htb]
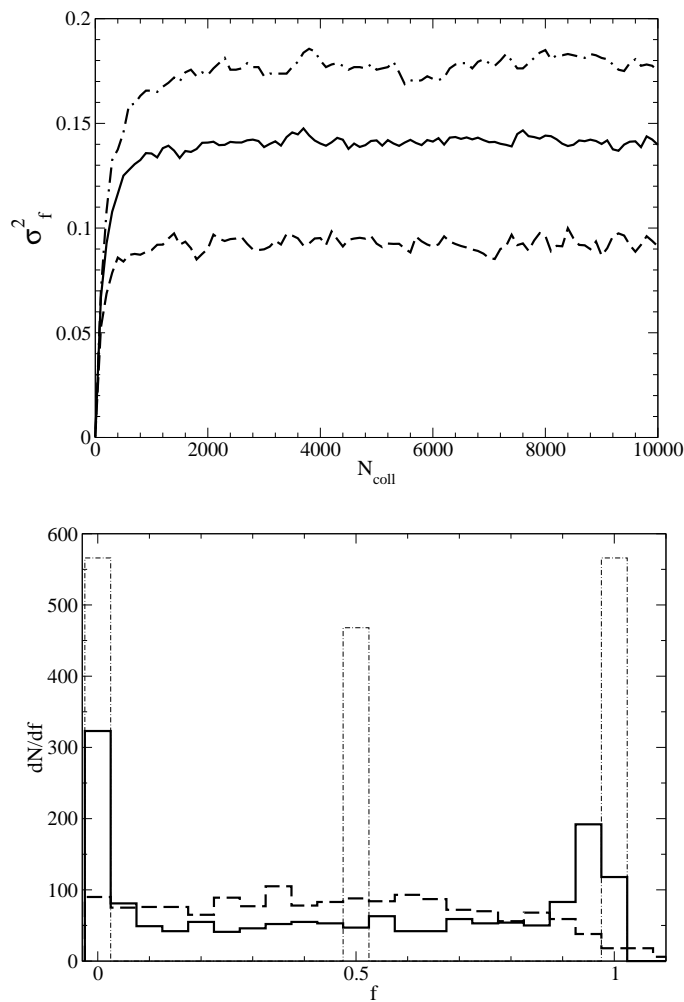

\begin{center}
\includegraphics[scale=0.31]{fig4a_flu.eps}
\end{center}
\vspace{0.5cm}
\begin{center}
\includegraphics[scale=0.31]{fig4b_flu.eps}
\end{center}
\caption{Upper panel: variance of the distribution function in $V_p$, as a function of the
number of collisions,  for three
different cases: fixed grid with $f(t=0) = 0.5$ (dot-dashed), fixed grid with randomly 
distributed test particles (full), moving grid (dashed). Lower panel:
Distribution of the number of cells as a function of their occupation $f$ at equilibrium,
for the three cases indicated above.}
\label{full_esa}
\end{figure}

The dot-dashed
curve in Fig.\ref{full_esa}, upper panel, shows the variance 
of the distribution function 
calculated
in the cells of the fixed grid, with increasing number of collisions
(i.e. as a function of time). 
It saturates 
approximately when almost all nucleons have suffered one collision. 
In the equilibrium configuration the number of 
cells with the three possible values of $f$ is roughly the same, as shown
 in Fig. \ref{full_esa} (lower panel), so the fluctuation variance is  
$\sigma^2_f\approx 0.175$, which is about $2/3$ of the maximum. 

The fluctuations reached in $V_p$ 
also depend on the variance of the initial configuration. 
If we initialize the system distributing the test particles 
randomly, the occupation in the cells has an average value
$\langle f\rangle=0.5$ and a width given by 
$\sigma_{f}(t=0)/ \langle f \rangle=\bar{N}^{-1/2}$, 
where $\bar{N}$ is the 
average number of test particles per cell (${\bar N} = 250$). 
The distribution function can 
assume any value between $0$ and $1$, 
as shown by the full curve in Fig. \ref{full_esa}, lower panel.
The initial continuous distribution of values of $f$ causes a reduction 
of the fluctuations in $V_p$ 
(full curve in Fig.~ (\ref{full_esa}), upper panel) with respect to the 
previous case.
As expected, the fact that $f$ varies continuously between $0$ and $1$ 
contributes to
enhance the smearing of the nucleon wave packet and hence to reduce
fluctuations in $V_p$.  

\subsection{The centroid degree of freedom}
\label{centroid}
Now we turn to a more general situation, relaxing the condition 
of the fixed grid. 
Hence here we would like to test the full procedure, described in Section 4,  that we will finally
follow in the full 3D case. 
The two reference frames used to make 
the collision (the one concerning the initial states and the one for the
final states) 
are now shifted along the $\phi$ axis, since the angle $\phi$ of the initial
and final states can assume any value between $0$ and $2\pi$,
according to the initial and final positions of the 
colliding test particles, $i_1$ and $i_2$.  

This carries two 
important consequences. First of all, even being the Pauli blocking 
checked for each collision in a particular frame,
the prescription $f\leq 1$ can be violated in another frame. However, we 
checked that this violation concerns only about $5\%$ of the cells. Second, 
the correlation between the cells representing the reconstructed nucleon, 
which is maximum in the search frame, will be systematically broken 
in another (shifted) reference frame, 
increasing the  nucleon delocalization 
and thus reducing fluctuations in $V_p$.

The dashed curve in 
Fig. \ref{full_esa} shows the fluctuations 
obtained in $V_p$
in this case. 
Since this feature is due to the arbitrary 
position of the centroid of the constructed nucleon with respect to a fixed 
reference frame, we will name it \emph{centroid effect}. 
We notice 
a fluctuation reduction of $\sim 60\%$
with respect to the fixed grid case (dash-dotted line in the same 
Figure). 

\subsection{Sensitivity to the dimension of the grid cell}

\begin{figure}[htbp]
\centering
\includegraphics[scale=0.3]{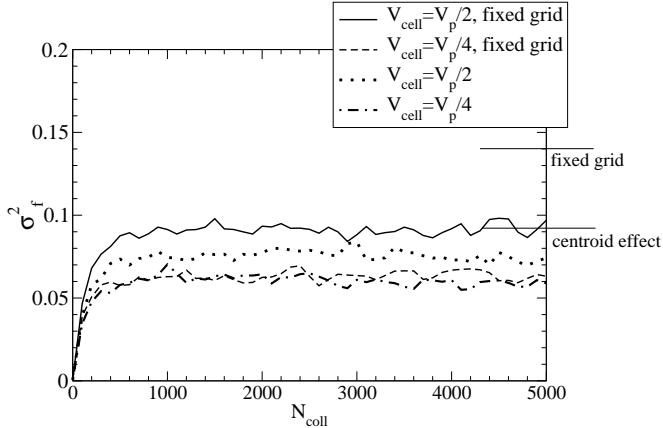}
\caption{Variance of the distribution function in the fixed grid scheme and in the 
general case, 
for two sizes of the search cell. The lines on the right indicate the values 
for $V_{cell}=V_p$, also shown in Fig. \ref{full_esa}.}
\label{smallphi_full}
\end{figure}

Up to now, the step of the grid has been chosen 
so that $V_{cell} = V_p$. 
However, in general, 
the size of the search cell
 has to be smaller than $V_p$, especially when the 
nucleon wave packet is rather delocalized in ${\mathbf p}$ space (being more
compact in coordinate space). 
In fact, small search cells are required for the 
Pauli blocking to be accurate. 
Hence we want 
to study the sensitivity of the results to  
the size of the search cell. 
Let us start with the fixed grid scheme.
In Fig. \ref{smallphi_full} we plot the 
fluctuation variance obtained 
dividing the step in the $\phi$ direction, $\phi_{step}$, 
by two or four; 
the volume of the search cell is then 
$V_{cell} = V_p/2, V_p/4$, respectively.  
This is equivalent to shift, for each collision, 
the origin of the grid by $0.5$ or $0.25$ $\phi_{step}$ 
respectively, and acts also 
as a sort of centroid effect, thus reducing 
the variance in $V_p$ (see solid and 
dashed lines in Fig. \ref{smallphi_full}). 
The value obtained with $V_{cell} = V_p/2$ is 
lower than the corresponding case with  $V_{cell} = V_p$,
indicated as ``fixed grid'' in the Figure.
Moreover, 
it further reduces for $V_{cell} = V_p/4$. 
Nevertheless, this smearing effect is partly compensated by the fact that the 
procedure used to select the cells belonging to the nucleon clouds 
favours compact, symmetric configurations.

It is interesting to evaluate the effect  
of the use of a small search cell on the fluctuations 
also in the general case, i.e. when the fixed grid
condition is relaxed. 
Intuitively, we expect 
a small effect on the variance, as long as $V_ {cell}$ is not too small: 
the main reduction of fluctuations comes 
already from the {\it centroid effect} (see previous Subsection), 
which is equivalent to
an infinitely small step. 
In fact, 
the differences between calculations with $V_{cell} = V_p/2$ 
and $V_{cell} = V_p/4$  (Fig. \ref{smallphi_full}, dotted and
dash-dotted curves respectively) are smaller than in the 
previous calculation with fixed grid.
The reduction with respect to the case with $V_{cell} = V_p$ (denoted as ``centroid effect'' in the Figure) 
is also smaller.


\subsection{Conclusions on the search procedure}
In conclusion, it appears from our calculations that 
the smearing of the nucleon wave packet 
is rather dependent on the procedure adopted to build nucleons and is
particularly affected by the so-called {\it centroid effect}, related to 
the use of randomly distributed test particles for the phase-space mapping. 
 As a consequence, fluctuations 
calculated in volumes $V_p$
appear reduced.
As shown above, 
the use of a fixed grid to locate the initial and final cells
of the colliding test particles 
would help 
to increase the value of the variance in $V_p$.  
However, 
energy and momentum conservation would not be exact, due to the
phase-space discretization. 
On the other hand, the standard test particle method allows a good mapping of the phase space and a 
good resolution of the average collision integral, with perfect energy and
momentum conservation. However, when constructing fluctuations, 
the {\it centroid effect}
may destroy the correlations already built and
increase the nucleon cloud smearing. 
However, the expected statistical value  
of the fluctuation variance 
is recovered
in large volumes.

\begin{figure}[htbp]
\vspace{0.8cm}
\begin{center}
\includegraphics[scale=0.3]{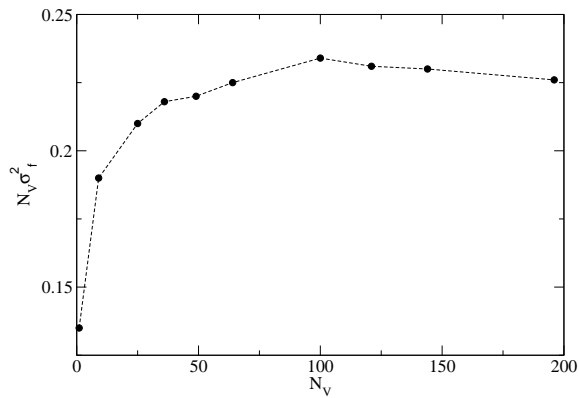}
\end{center}
\caption{Variance of the distribution function (rescaled by $N_V$) as a 
function of the number $N_V$ of nucleon cells $V_p$, obtained 
in the fixed grid case with $V_{cell}=V_p$ and test particles randomly distributed.}
\label{fluct_new}
\end{figure}

This last point has been checked 
for all the simplified situations considered here (Fermi surface). 
For instance, this analysis is shown in Fig.\ref{fluct_new} in the case 
of the fixed grid with $V_{cell} = V_p$ and test particles randomly distributed
inside the system (last case of Section 5.2).
Here fluctuations are evaluated in volumes $V$, by considering
an ensemble of 800 events. 
As expected, the fluctuation variance, rescaled by $N_V$,  almost approaches
the expected value as the volume $V$ increases. 
As explained above for the idealized case presented in Fig.3, 
the maximum value, $N_V\sigma^2_0 = 0.25$, is not exactly reached 
because of the upper bound imposed to the correlation 
radius in the nucleon reconstruction procedure and of mass conservation constraints.


\section{Non-equilibrium situations}\label{init_distr}

In the Fermi energy domain, two colliding 
nuclei may be schematically represented 
by touching Fermi spheres in $\mathbf{p}$ space, so
it is interesting to study the 
development of fluctuations starting from 
this initial distribution. 

In our simplified 2D scheme, we can mimic the two spheres using  a ``chess-board'' initialization, that is 
subdividing our momentum space in four regions where $f$ is alternatively zero or one. 
Although the system is in a non-equilibrium situation, the variance, 
defined as in Eq.(\ref{sigmaf}), starts 
from its maximum value 
(since $\langle f \rangle =0.5$). 
It can be shown that in the fixed grid case 
the equilibrium configuration consists of a random 
distribution of full and empty cells, and the variance keeps constant in time. 
However, a very different result comes out either changing 
the search step or removing the fixed grid constraint. In both simulations 
the variance rapidly decreases because the initial correlation is destroyed 
by the {\it centroid effect}. 

\begin{figure}[htbp]
\vspace{0.8cm}
\begin{center}
\includegraphics[scale=0.3]{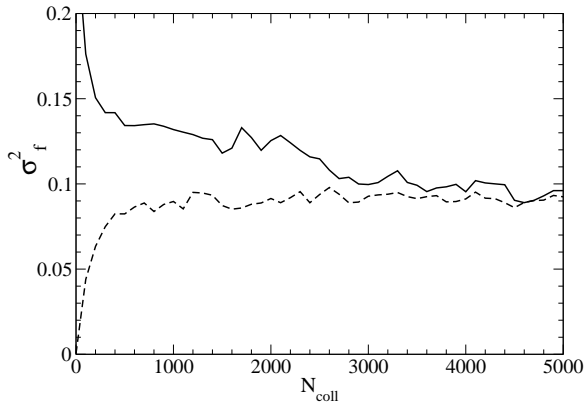}
\end{center}
\caption{Comparison between variances obtained with the ``chess-board''
initialization (solid line) and with the case
$f = 0.5 \pm\sigma_{f}(t=0)$ 
(dashed curve) in the fixed grid case for $V_{cell}=V_p/2$.}
\label{f1_full}
\end{figure}

\begin{figure}[htbp]
\vspace{0.8cm}
\begin{center}
\includegraphics[scale=0.3]{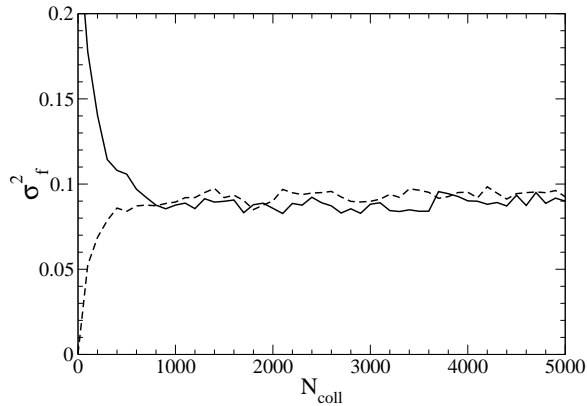}
\end{center}
\caption{Comparison between variances obtained with the same initializations 
of Fig. \ref{f1_full} but for $V_{cell}=V_p$ and 
without the fixed grid constraint.}
\label{f1_rand}
\end{figure}

As an illustrative example, in Fig. \ref{f1_full} 
we plot $\sigma^2_f$ for the choice $V_{cell}=V_p/2$, fixed grid. 
Comparison with the corresponding curve
for the configuration $f(\mathbf{p})=0.5 \pm\sigma_{f}(t=0)$ 
(dashed, see Section 5.4) 
shows that the two calculations converge to the same value.  
We get a confirmation of these findings 
by removing the fixed grid constraint. 
Also in this case, at the end we find no difference with  the case described 
before (initialization with randomly distributed test particles, Section 5.3), see Fig. \ref{f1_rand}.
Moreover, in this case the convergence between the two calculations is rather quick.
Then we can conclude that 
the final outcome does not depend on the initial 
conditions of the system, but only on the final equilibrium situation, as it should be. 


\section{Results for full 3D simulations} 


Guided by the simplified schematic cases illustrated above, 
let us now discuss the full case of a nuclear Fermi gas of particles at a given
temperature and density interacting through hard two-body scattering.  

As in the previous examples, 
since we focus on fluctuations in momentum space, only one large cubic cell is 
present in coordinate space, and all particles can be chosen to collide 
(no restrictions in $\mathbf{r}$ space). 
The size of the box is $L=26\,fm$, and we consider $2820$ nucleons, so that 
the density has the saturation value $\varrho_0=0.16\,fm^{-3}$; each nucleon 
is represented by a collection of $500$ test particles. 
 Besides, we do not consider 
any distinction between neutrons and protons, so that one nucleon 
occupies at least a phase-space volume $h^3/4$ ($g = 4$). 
We initialize the momenta so as to 
reproduce a Fermi-Dirac profile corresponding to a temperature of $5\,MeV$. 
Finally, we consider a constant cross section $\sigma=160\,mb$. 
In these calculations the volume $V_{cell}$ corresponds to a cube of side  
$l_s=30\,MeV/c$ (and coincides with $V_p = h^3/(4L^3)$).
We notice that, for the full Fermi gas case, a grid
in Cartesian coordinates is easier to use, with respect to
spherical coordinates, in the nucleon construction procedure. 
However, spherical coordinates will be adopted to analyze the resulting
fluctuations. 
Calculations are stopped when the fluctuation variance saturates.  

\begin{figure}[htb]
\vspace{0.4cm}
\centering
\includegraphics[scale=0.25]{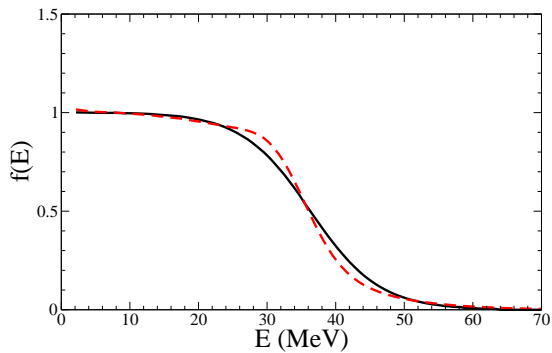} 
\caption{Distribution function profile at the initial time $t=0$ (solid) 
and at the time $t=100\,fm/c$ (dashed line).}
\label{effe}
\end{figure}

First of all, we checked 
the effect of our fluctuating collision integral on
the average evolution of the system. 
In Fig. \ref{effe} we plot the energy 
profile of the distribution function at the initial time (solid) and 
at the time when the fluctuation variance saturates and 
we stop the calculation (dashed line). Slight changes are 
observed, probably due to 
the finite extension of the nucleon packet, 
that induces a discretization of the phase space.

We also compared the average collision rate with analytical expectations: 
we find that the number of collisions is slightly larger than the 
analytical one, but differences are within $10\%$; on the other 
hand, similar deviations are also found in usual BUU implementations, 
due to the approximate mapping of phase space 
induced by the finite number of test particles.

\begin{figure}[htb]
\centering
\includegraphics[scale=0.34]{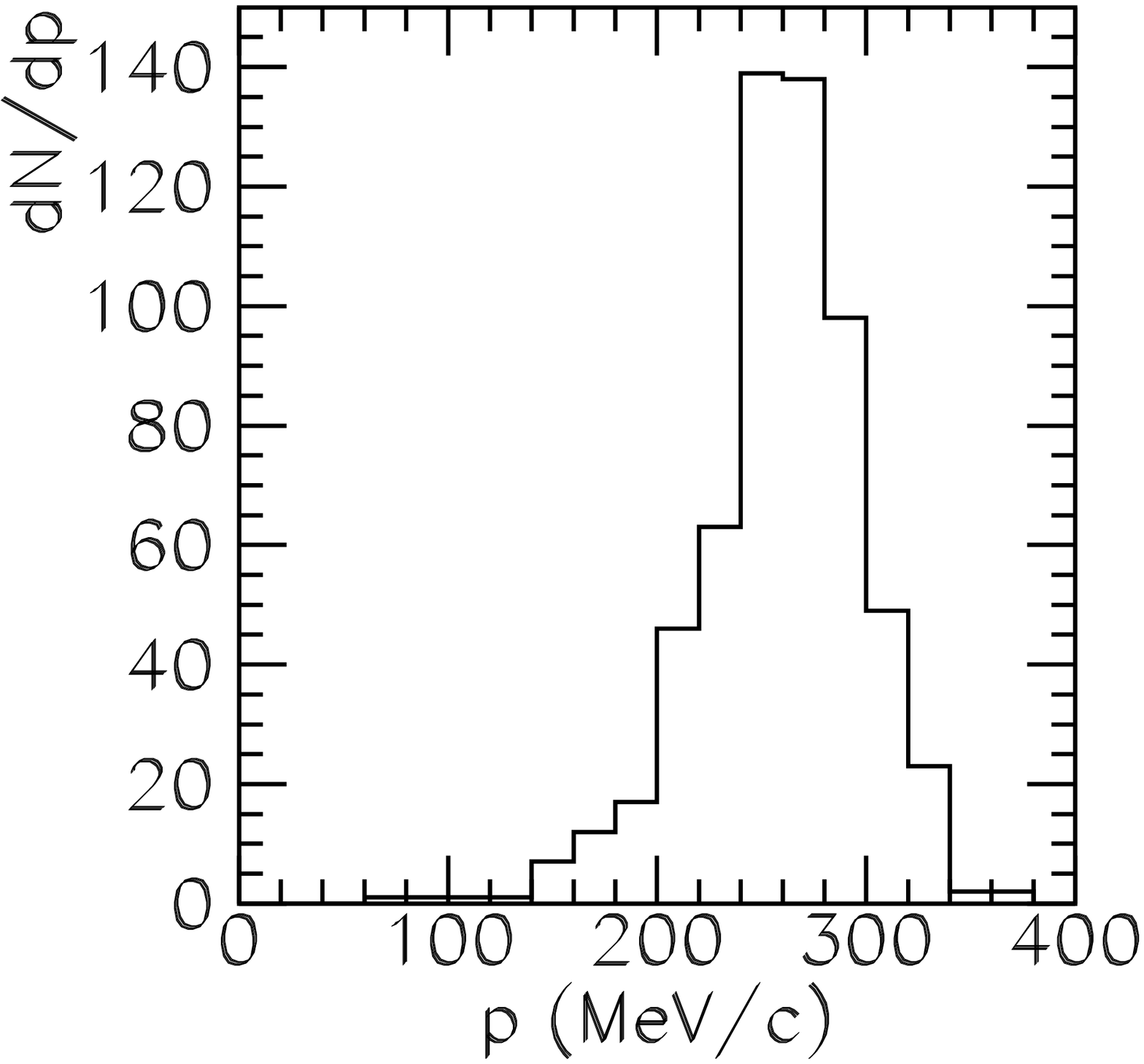}
\includegraphics[scale=0.34]{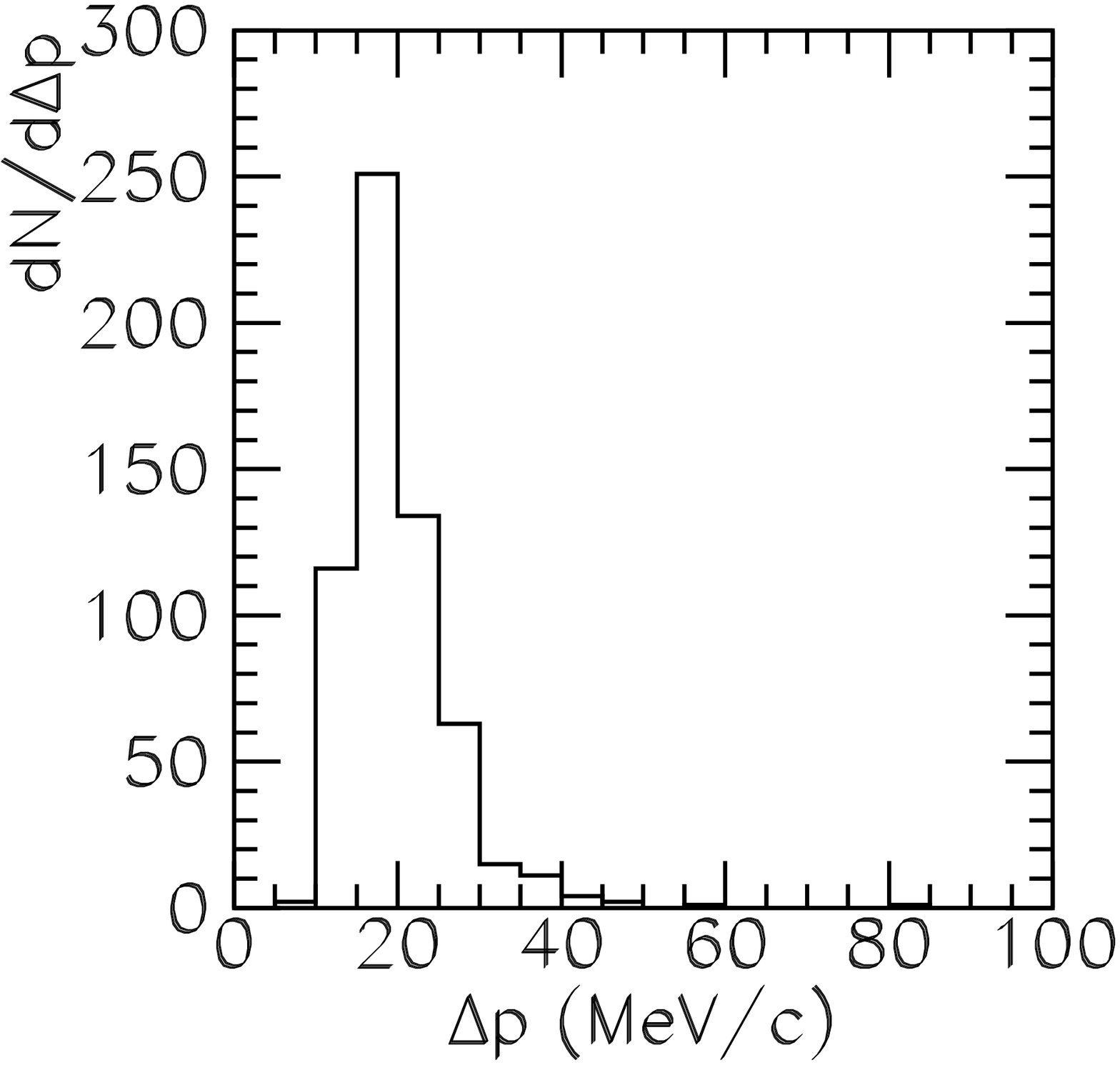}
\caption{Left panel: distribution of the position of the centroid of the 
nucleon ``clouds'' in the $|p|$ direction. Right panel: distribution of 
the size of the nucleons on the same axis.}
\label{polar}
\end{figure}

In the following  we analyze in more detail the capability of the procedure to build the 
fluctuations in the considered 3D scenario.
We investigated the region in ${\bf p}$ space
where collisions 
are most probable and the extension of the nucleon cloud.  
From the left panel of Fig. \ref{polar} it is evident 
that most collisions involve nucleons lying on the surface of the Fermi 
sphere, as expected. Besides, the right panel illustrates the extension of 
the clouds in the radial direction, calculated according to the definition:
\begin{equation}\label{delta_p}
\Delta p =\sqrt{\langle(p_i-\langle p\rangle)^2\rangle_{cloud}}, \quad 
\end{equation}
where $\langle p\rangle$ denotes the center of momentum of the cloud and
$p_i$ is the modulus of the momentum of the test particles contained in the cloud. 
 
\begin{figure}[htb]
\vspace{1.2cm}
\centering
\includegraphics[scale=0.3]{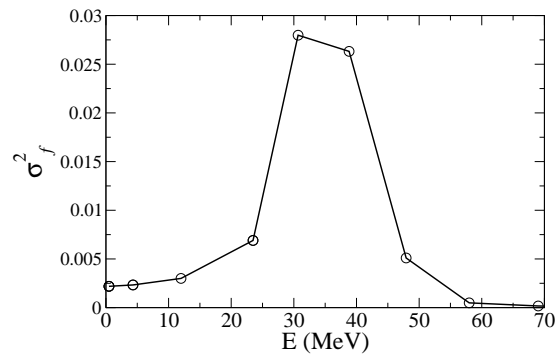}
\caption{Profile of the fluctuation variance, as a function of the 
energy $E = p^2/(2m)$, in cubic cells
of size $V_p$, for a Fermi gas at equilibrium.}
\label{flu_vp}
\end{figure}

\begin{figure}[htb]
\vspace{0.4cm}
\centering
\includegraphics[scale=0.31]{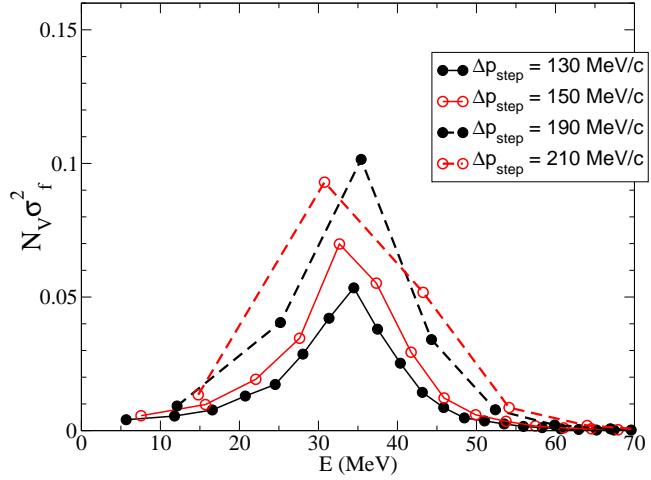} 
\caption{Fluctuation variance as a function of the energy, for a fixed angular spread 
$\theta=30^{\circ}$ and different steps $\Delta p_{step} = (\Delta p^3)^{1/3}$.} 
\label{spicchi_p}
\end{figure}

From this distribution one can deduce that the clouds 
correspond to a relatively narrow stripe, $2\Delta p \approx 35 MeV/c$, 
in the region where $f$ is different 
from either zero or one. 
Moreover, we have checked that the extension of the constructed wave packet  
is similar in the three directions $p_x$,$p_y$,$p_z$,  
as expected. 
We notice also that the tail  of the distribution (Fig.10, right)
extends up to large $\Delta p$ values, $\approx 50 MeV/c$.
Although the nucleon clouds extends, on average,
over a volume $(2\Delta p)^3 \sim~(35\,MeV/c)^3$,
that is not much larger than $V_p$, the {\it centroid effect}, i.e. the fact that
the nucleon centroid may be located anywhere, 
causes a significant smearing of fluctuations, as already discussed in 
Section 5.
In fact, we find that, for our Fermi gas calculations, fluctuations
in $V_p$ are considerably underestimated, as shown in Fig.\ref{flu_vp},
where the variance $\sigma^2_f$ is reported as a function of the energy
$E  = p^2 / (2m)$. 
In the following, to test our procedure,   
we will evaluate fluctuations considering big volumes, 
that contain many nucleons. 
However, it should be noticed that, in the geometry considered, 
it is not trivial to predict which is the size and shape of 
the most appropriate volume to recover the expected value of fluctuations. 
While, from one side, it is desirable to consider volumes containing
a large number of nucleons, from the other side one should keep in mind that
spurious fluctuation reduction, due to the 
conservation of the total number of particles, 
may be a problem.
Hence we have performed a systematic analysis of fluctuations 
as a function 
of the coordinates 
$(p,\theta)$, integrating over the angle $\phi$. The 
corresponding volume employed for the calculation of the variance is 
given by:
$$
V=2\pi \frac{\Delta p^3}{3}\Delta \cos\theta
$$
We consider volumes corresponding to fixed angular spreads 
$\theta$
and/or to a given step $\Delta p^3$. 
In Fig. \ref{spicchi_p} we fix the angular spread, $\theta=30^{\circ}$,
and we consider four 
different steps, $\Delta p^3$, 
corresponding to different volumes, i.e. to a
different number of nucleons $N_{V}$ that can be contained inside.
The considered volumes can
contain up to a rather large number of nucleons (about 90). 
The variance, multiplied by $N_V$,  is plotted  
as a function of the energy $E$. 
We find 
that the peak of the variance is around the Fermi energy, as expected.  
The rescaled variance gets larger when we increase the step $\Delta p^3$.
However, even in the best situation, the expected equilibrium 
value at the peak ($N_V\sigma_0^2 = 0.25$) is underestimated by a factor 
$\sim 2.3$. 
Therefore, for a fixed value of $\Delta p_{step} = (\Delta p^3)^{1/3}
=190\,MeV/c$, we tested different
angular spreads. 
\begin{figure}[htb]

\centering
\includegraphics[scale=0.3]{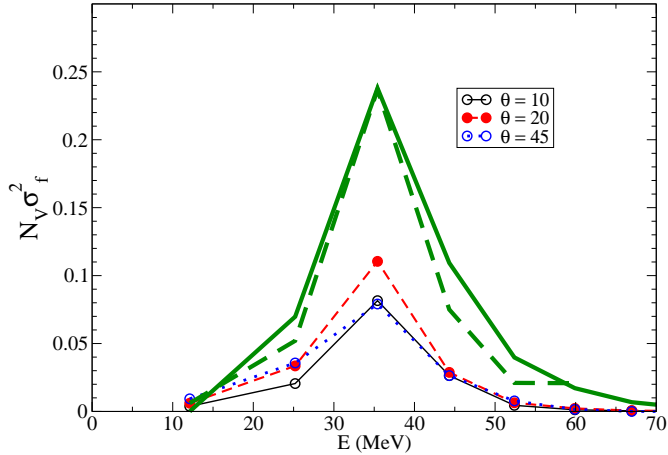}
\caption{Fluctuation variance as a function of the  energy, for a fixed $\Delta p_{step}=190\,MeV/c$ 
and different angular spreads (curves with circles). The thick full and dashed
lines are explained in the text.} 
\label{spicchi_teta}
\end{figure} 
In Fig.\ref{spicchi_teta} we present the obtained results, rescaled by
$N_{V}$.
Now we can observe that, increasing the angular spread, the 
variance first increases, then decreases again, due to the mass
conservation constraints.
The variance gets larger 
when considering the angular spread $\theta=20 ^{\circ}$, but the expected value is still
underestimated by about a factor $2$, at the peak. 
This result can be understood on the basis of
the results obtained in $V_p$ and displayed 
in Fig.\ref{flu_vp}. 
In fact, 
in order to recover  the expected fluctuation value, 
the volume employed for the calculation of the variance must be much 
larger than $V_p^{1/3}$ 
in all the directions $(p,\cos (\theta),\phi)$. 
Due to the considered geometry, 
while this  
can be achieved in the angular coordinates, this cannot be easily obtained 
along the $p$ direction without losing locality for the value
of $f$ and related fluctuations.
Actually,  in the case of $\Delta p_{step} = 190\,MeV/c$, 
the corresponding extension along the $p$ direction,
$\Delta p_{mod}$, 
of the volume considered to evaluate the function
$f$ is approximately equal to $30 MeV/c = V_p^{1/3}$ at $p=p_F$. 
We thus
expect fluctuations to be reduced, with respect to $f_{eq}{\bar f}_{eq}$, 
by a factor 
$\alpha^{1/3}\approx 0.48$, being $\alpha\approx 0.11$ the
reduction factor in $V_p$, as evaluated from Fig.\ref{flu_vp}. 
The expected fluctuation reduction can be extracted also from 
the schematic calculations described in Sect.5. 
In fact, 
from the inspection of Fig.\ref{full_esa}, it appears that, in the
fixed grid scheme, fluctuations in $V_p$ are reduced by a factor $0.56$.
Relaxing the fixed grid constraint in one 
dimension, fluctuations are further reduced
by a factor 0.64 with respect to the fixed grid case. 
We thus
expect a fluctuation reduction 
by a factor $0.56^{1/3}\times 0.64 \approx 0.5$ (at the Fermi surface).
This is close to our result at $E=E_F = p_F^2 / (2m)$
(see the dashed line with circles in Fig.13, for instance).

One may try to introduce this correction in the results of Fig. 13.
In the considered geometry, 
the reduction factor $\alpha$ of the fluctuations in $V_p$ depends 
on the energy $E$, i.e. one can write, for the variance in $V_p$:
\begin{equation}
\label{sig1}
\sigma^2_f(E,N_V=1) = {F}(E)~\alpha(E),
\end{equation}
where  ${F}(E)$ indicates the expected fluctuation value, that
in our case should coincide with $f_{eq}{\bar f}_{eq}$.

By approximating the volume dependence of the rescaled variance 
by a linear behaviour for small volumes (see Fig.\ref{flu_joseph} for instance), 
the fluctuation variance presented in Figs. 12-13 can be rewritten as:
\begin{equation}
\label{sig2}
N_V \sigma^2_f(E,N_V) \approx {F}(E)~ \alpha^{1/3}(E)
\frac{\Delta p_{mod}}{V_p^{1/3}}.
\end{equation}
We notice that $\Delta p_{mod}$ is also depending on $E$. 

Eqs.(\ref{sig1})(\ref{sig2}) allow to extract the suppression
factor $\alpha(E)$, as well as the function ${F}(E)$. 
The latter is displayed in Fig.13 
(thick dashed line).  One can see that 
the expected fluctuation value, $f_{eq}{\bar f}_{eq}$ (thick solid line),  
is well reproduced at the Fermi surface, while it is underestimated, within $30\%$,
expecially for outer regions.
This underestimation indicates that 
in our procedure, 
due to the finite extension of the nucleon cloud, the regions far 
from the Fermi surface 
are less involved in the building of fluctuations than they should be
and can be considered as an intrinsic limitation of the method. 

\begin{figure}[htb]
\centering
\includegraphics[scale=0.36]{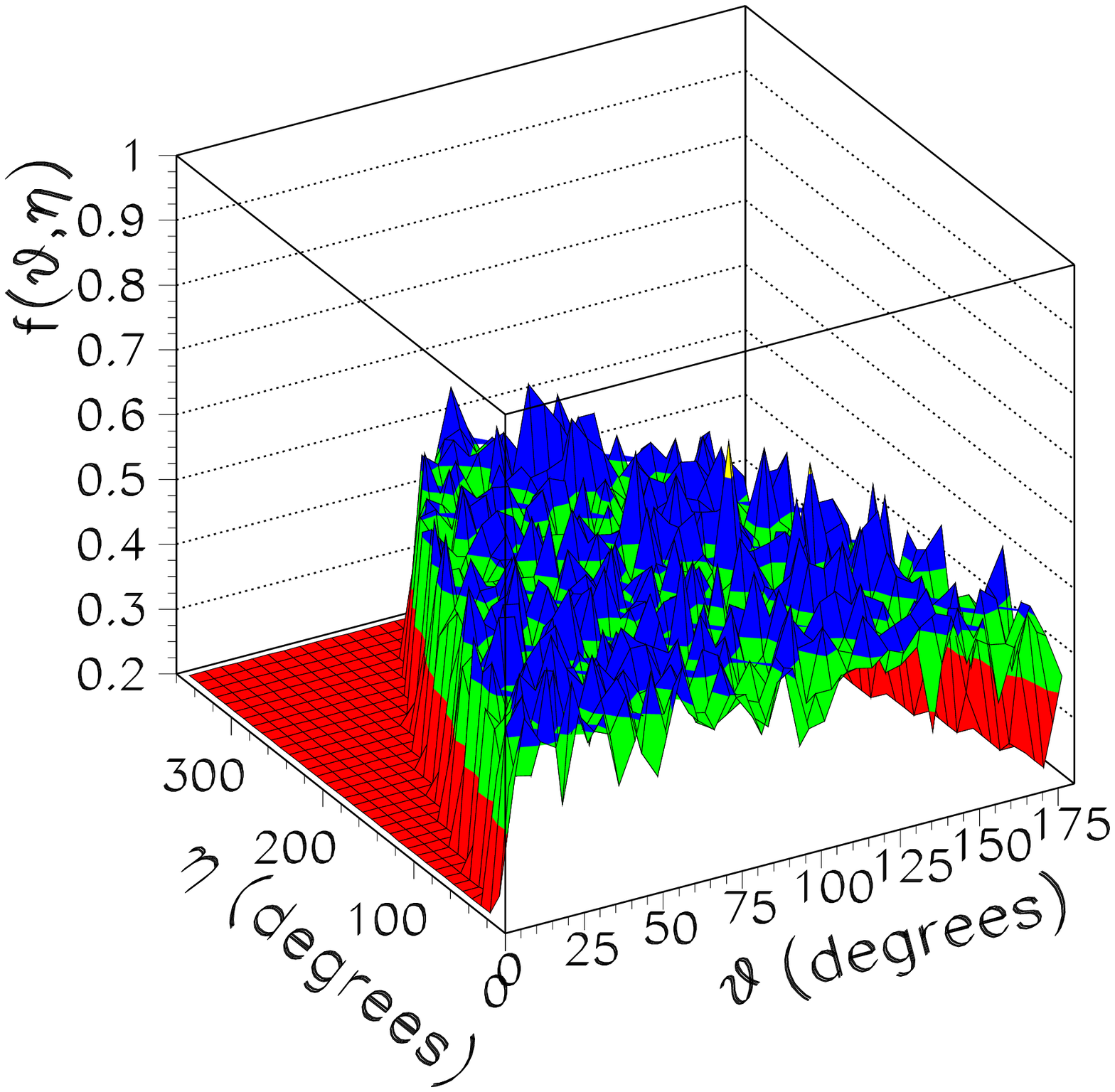}  
\includegraphics[scale=0.36]{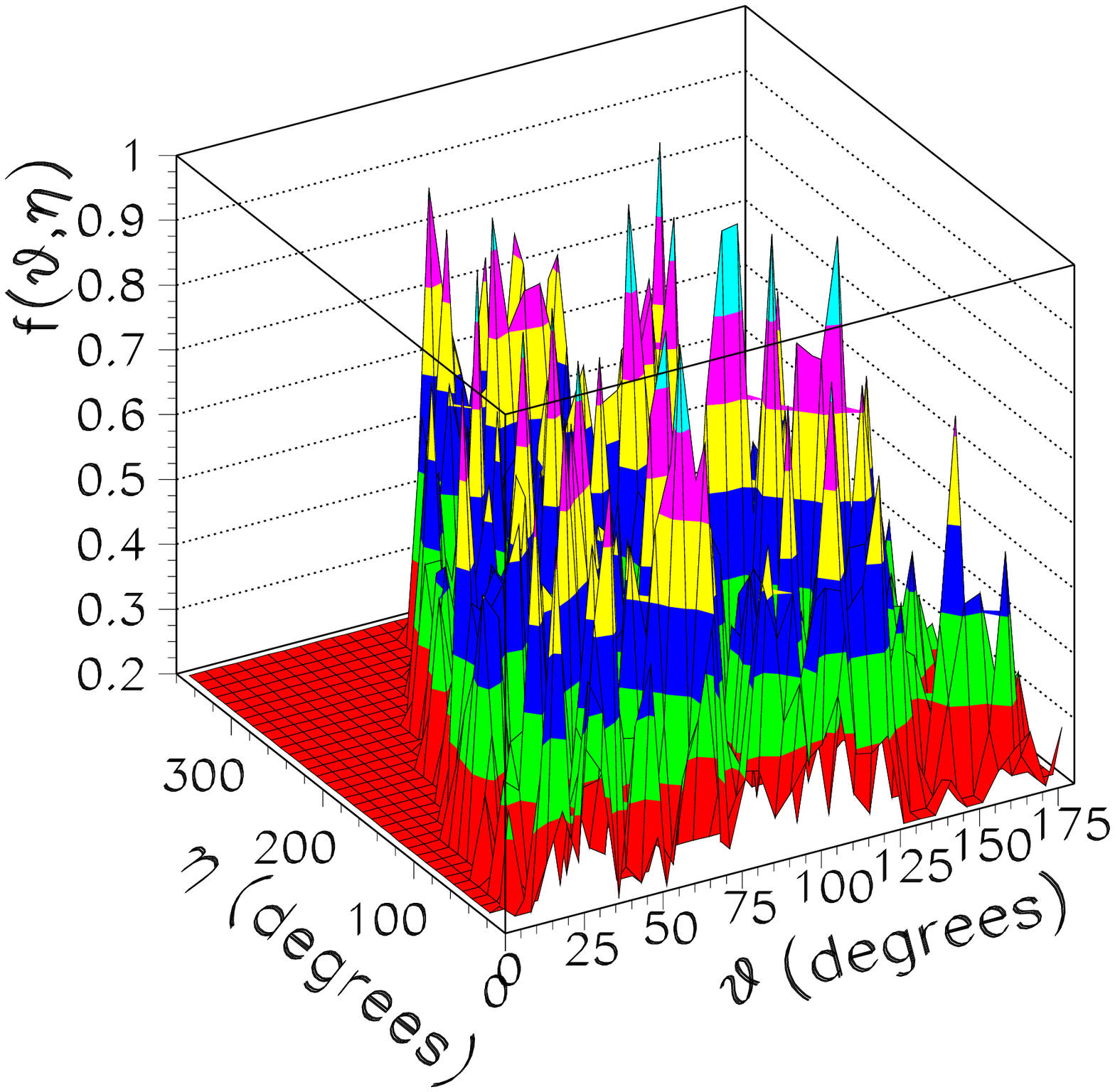}  
\caption{Distribution function $f(\theta,\eta)$ at time $t=0$ (left) 
and $t=100\,fm/c$ (right).}
\label{tetaeta}
\end{figure}

However, 
the overall shape of the fluctuation variance is reasonably reproduced, relative to
$f{\bar f}$:  
This is a nice indication that the Pauli blocking is 
effective, and the equilibrium distribution is almost unchanged by the 
procedure. Therefore, it really represents a remarkable improvement with 
respect to the original method by Bauer et al. (see Ref.\cite{chapelle}), 
which makes us confident 
about applications to nuclear reactions. 
We can directly visualize the 
effect of the fluctuating term by selecting a thin region in momentum space 
around the Fermi momentum. 
For this purpose, we adopt 
a new set of coordinates $(dp,\, pd\theta,\, p \sin(\theta)d\phi )$;  
for fixed $p$ and $dp$, the distribution function depends only on 
two coordinates, namely: 
$$
f(p,\theta,\phi)\rightarrow f(\theta,\eta)
$$
with $\eta=\sin(\theta)\phi$. 
This set of coordinates correspond to displacements of equal length
along the three directions: $p,\theta,\phi$.
We choose $p=260\,MeV/c$ (approximately equal to the Fermi momentum) and 
$dp=10\,MeV/c$, and in Fig. \ref{tetaeta} we plot the distribution 
function $f(\theta,\eta)$ at two different times (initial and final times). 
We recognize the sinusoidal 
profile on the $\eta$ axis due to the modulation given by the $\sin(\theta)$ 
factor. At the initial time, $f$ is nearly uniform, and its fluctuations are 
simply due to the numerical noise induced by the finite number of test 
particles. At later times, we observe 
the growth of fluctuations, evidenced by the typical structure with ``peaks 
and holes''. 


\section{Methods to increase fluctuations in $V_p$}

\label{optimize}

In actual nuclear collisions the extension of the nucleon wave packet
in phase-space may affect the transport dynamics and it may be important
to correlate the test particles belonging to the nucleon clouds 
inside smaller volumes.   
In the context of the simplified model (Fermi surface) discussed in Sect.5
we attempted to find a method to 
reduce the nucleon cloud smearing and 
enhance the fluctuations in $V_p$ without losing 
the consistency of the original procedure. 

First of all, it should be remembered that the collision probability 
is given by $f(\mathbf{p}_1) \bar{f}(\mathbf{p}_3)$ 
(see Sect. \ref{details}): therefore, we need to consider only 
two locations in 
$\mathbf{p}$ space. 
The main idea is to force the system to fully empty the cells associated with the initial 
state or fully fill the final state when a collision occurs.
This condition is not automatically achieved for all cells contributing to 
a given collision.
In fact, 
the number of particles, $n_t(i)$, 
taken from a cell $i$ and employed to build the 
nucleon obey to the constraint: 
$$
\sum_i n_t(i)=N_{test}
$$
being $N_{test}$ the number of test particles per nucleon. 
Hence the number of test particles that are really taken from a cell
is not always 
equal to $n_t=min(n_1,\bar{n}_3)$, but instead to $n'_t = min(n_t,n_{res})$,
being 
\begin{equation}
n_{res}(i)=N_{test}-\sum_{i'=1}^{i-1}n_t(i')
\label{prob2}
\end{equation}
The idea is to try  to
avoid the situation when the number of particles moved 
is less than $n_t$.
In this way 
fluctuations get to 
a larger value, since 
we favour the transitions where cells are completely
filled or emptied. 
In order to implement this idea, when constructing the nucleon clouds  
 we choose, among the possible ones, the cells for which the 
probability $p_t = \frac{n'_t}{n_t}$ is large. 

\begin{figure}[htb]
\begin{center}
\includegraphics[scale=0.25]{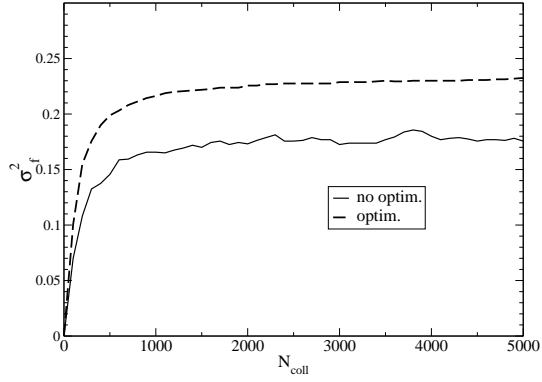}
\end{center}
\caption{Variance of the distribution function in $V_p$,
as a function of the number of collisions, 
in the fixed grid scheme.
Solid line: no optimization; 
dashed line: optimized procedure. 
}
\label{opt_full_esa}
\end{figure}
The outcome of this method is plotted 
in Fig. \ref{opt_full_esa} for the fixed grid case, where the result 
of Fig. \ref{full_esa} 
is also shown for comparison (solid line). 
As the dashed curve in Fig. \ref{opt_full_esa} clearly shows, 
this optimization procedure is successful in 
enhancing the fluctuations, which eventually 
reach almost their upper limit $\sigma^2_0=0.25$. Actually, 
this maximum value is not obtained 
due to the fact that  only  
adjacent cells are considered in the nucleon construction. In this way, some 
half-filled cells remain ``isolated'' and 
are not involved in a 
collision anymore. Indeed, at the time 
when we stop the calculation, isolated cells 
amount to about $5\%$ of the total. 


The method works also in more general situations, for instance 
in the case corresponding to fluctuating initial conditions (see Sect.5.2).
\begin{figure}[thbp]
\begin{center}
\includegraphics[scale=0.25]{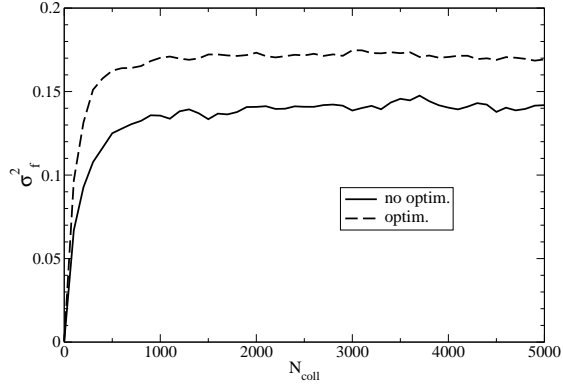}
\end{center}
\caption{Variance of the distribution function in $V_p$, as a function of the
number of collisions, in the fixed grid scheme 
with fluctuating initial conditions. 
Solid line: no optimization; dashed line: optimized procedure. 
}
\label{opt_full_rand}
\end{figure}
The comparison between variances with and without 
the optimization procedure is shown in Fig. \ref{opt_full_rand}. As in the 
previous case, the optimization favours a faster growth of the fluctuations. 
Their saturation value for $V=V_{p}$ is 
now enhanced, by $\sim 20\%$ with respect to the standard procedure. 
This is an indication that the procedure is able to 
increase the number of empty and full cells. This feature is also evident  
from the distribution of values of $f$ plotted,  for the same 
two simulations of Fig.16, in Fig. \ref{opt_dndf_full}.

In the end, this procedure helps the 
nucleon packets to keep a similar, more compact shape in $\mathbf{p}$ space during 
the time evolution of the system, from one collisional event to another.
As a consequence, 
nucleons are more localized and 
fluctuations in $V_p$ appear enhanced.  

\begin{figure}[tbp]
\vspace{0.3cm}
\centering
\includegraphics[scale=0.25]{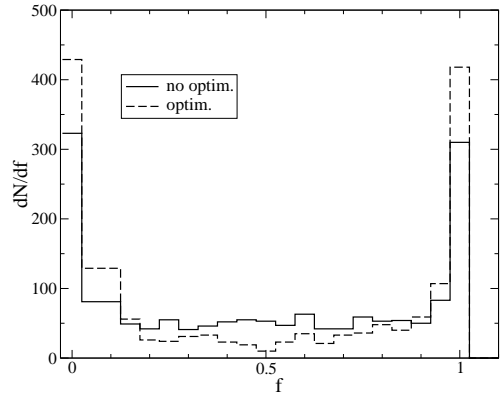}
\caption{Distribution of the number of cells as a function of their occupation $f$ for the same 
two calculations of Fig. \ref{opt_full_rand}.}
\label{opt_dndf_full}
\end{figure}

\section{Conclusions}
We have proposed a new numerical implementation of the full Boltzmann-Langevin
equation in 3D. The stochastic character of the two-body collision integral,
that in standard transport codes applies only to the single test
particles, leading to a strong reduction of fluctuations,  
is recovered, at  the nucleon level, by a careful reconstruction 
of the nucleon wave packet once collisions occur.
This is achieved by moving, in each collisional event, two entire test particle clouds, 
corresponding to one nucleon each, as proposed in Ref.\cite{fluct_bauer}.
However, while in the original procedure the Pauli blocking was checked
only for the centroids of the nucleon clouds, here we propose a new method that 
allows to construct nucleon 
wave packets by carefully checking
the Pauli blocking for the entire swarm of test particles.
The last point is essential in the construction of the correct fluctuation
value in fermionic systems. 
The nucleon wave packet may take, in principle, any shape. Thus the correlation volume, i.e. the volume of the
sphere that contains all test particles that move together 
is generally larger than $h^3$. Moreover, the nucleon centroid is chosen 
randomly among the test particle distribution, leading to a partial
overlap of nucleon configurations from one collision to another. 
All these effects lead to a reduction of fluctuations in the cells 
of volume $h^3$.
However, the correct fluctuation value is recovered in larger volumes, 
where all possible nucleon configurations may be accommodated. 
 
The procedure, and its numerical ingredients, has been tested carefully 
in several idealized situations, studying in particular the relation 
between the results obtained for the fluctuations in cells of volume $h^3$ 
and the ingredients
of the model. 
Considering a fermionic system at equilibrium, at a given density and
temperature, we find that 
our procedure builds, within a good approximation,  the expected profile of the fluctuation 
variance,  $f_{eq}{\bar f_{eq}}$,  as
a function of the nucleon energy.
This can be considered as a significant improvement with respect
to the original procedure \cite{fluct_bauer}, that opens interesting possibilities
of applications to nuclear collisions.  
In fact, 
the method proposed can be easily implemented into existing transport
codes, and could allow to treat, in a more complete way,
phenomena where large fluctuations or bifurcations occur and
fluctuations are important.    

Finally we stress that the results presented here  are  relevant not only for nuclear
fragmentation studies, but in general for the dynamical description of 
quantum many-body systems. 

~\\[2ex]

\noindent
{\bf Acknowledgments}
\noindent
We warmly thank H.H.Wolter for the reading of the manuscript and 
stimulating discussions. 



\end{document}